\documentclass[english]{revtex4}
\usepackage[T1]{fontenc}
\usepackage[latin9]{inputenc}
\usepackage{color}
\usepackage{graphicx}
\usepackage{esint}

\makeatletter
\@ifundefined{showcaptionsetup}{}{%
 \PassOptionsToPackage{caption=false}{subfig}}
\usepackage{subfig}
\makeatother

\usepackage{babel}
\begin{document}

\title{Complex spectrum of finite-density lattice QCD with static quarks
at strong coupling}

\author{Hiromichi Nishimura}
\email{hnishimura@physics.wustl.edu}
\affiliation{Institute for Theoretical Physics, Johann Wolfgang Goethe-University
Max-von-Laue-Str.~1, D-60438 Frankfurt am Main, Germany}
\affiliation{Department of Physics, Washington University, St.~Louis, MO 63130
USA}
\author{Michael C. Ogilvie}
%\affiliation{Department of Physics, Washington University, St.~Louis, MO 63130 USA}
\email{mco@physics.wustl.edu}
\author{Kamal Pangeni}
\email{kamalpangeni@wustl.edu}
\affiliation{Department of Physics, Washington University, St.~Louis, MO 63130
USA}

%\address{Department of Physics, Washington University, St.~Louis, MO 63130 USA}

\date{1/20/2016}
\begin{abstract}
We calculate the spectrum of transfer matrix eigenvalues associated
with Polyakov loops in finite-density lattice QCD with static quarks.
These eigenvalues determine the spatial behavior of Polyakov loop
correlations functions.  Our results are valid for all values of the
gauge coupling in $1+1$ dimensions, and valid in the strong-coupling
region for any number of dimensions. When the quark chemical potential
$\mu$ is nonzero, the spatial transfer matrix $T_s$ is non-Hermitian.
The appearance of complex eigenvalues in $T_s$ is a manifestation of
the sign problem in finite-density QCD. The invariance of finite-density
QCD under the combined action of charge conjugation $\mathcal{C}$
and complex conjugation $\mathcal{K}$ implies that the eigenvalues
of $T_s$ are either real or part of a complex pair. Calculation of
the spectrum confirms the existence of complex pairs in much of the
temperature-chemical potential plane. Many features of the spectrum
for static quarks are determined by a particle-hole symmetry. For
$\mu$ small compared to the quark mass $M$, we typically find real
eigenvalues for the lowest lying states. At somewhat larger values
of $\mu,$ pairs of eigenvalues may form complex-conjugate pairs,
leading to damped oscillatory behavior in Polyakov loop correlation
functions. However, near $\mu=M$, the low-lying spectrum becomes
real again. This is a direct consequence of the approximate particle-hole
symmetry at $\mu=M$ for heavy quarks. This behavior of the eigenvalues
should be observable in lattice simulations and can be used as a test
of lattice algorithms. Our results provide independent confirmation
of results we have previously obtained in PNJL models using complex
saddle points. 
\end{abstract}
\maketitle

\section{Introduction}

Although lattice simulations have given excellent first-principles
results for many observables of finite-temperature QCD, there has
been less clear success when the chemical potential $\mu$ is nonzero.
Finite-density QCD is one of the class of theoretical models that
has a sign problem: the partition function is a sum over complex weights
which cannot be interpreted as relative probabilities \cite{deForcrand:2010ys,Gupta:2011ma,Aarts:2013bla}.
Many methods have been used in attempts to overcome the sign problem
in finite-density QCD. Two methods that have received significant
attention recently are the complex Langevin technique \cite{Seiler:2012wz,Aarts:2013uxa,Sexty:2013ica,Sexty:2014zya,Aarts:2014bwa,Nagata:2015uga}
and the Lefschetz thimble approach \cite{Cristoforetti:2012su,Fujii:2013sra,Cristoforetti:2013qaa,Cristoforetti:2013wha,Mukherjee:2013aga,Cristoforetti:2014gsa,Aarts:2014nxa,Tanizaki:2015pua,Fujii:2015bua}.
We have recently explored the implications of complex saddle points
in phenomenological models of QCD at finite temperature and density
\cite{Nishimura:2014rxa,Nishimura:2014kla}. These models postulate
effective potentials for the Polyakov loop $Tr_{F}P_{x}$ and other
order parameters in such a way that the confinement-deconfinement transition
of quenched QCD is incorporated. 
The most realistic of these models are Polyakov-Nambu-Jona
Lasinio (PNJL) models, and include the effects of chiral symmetry
restoration \cite{Fukushima:2003fw}. In all the cases studied in \cite{Nishimura:2014rxa,Nishimura:2014kla}, 
a nonzero $\mu$ resulted in a complex
saddle point for the eigenvalues of the Polyakov loop. A number of
desirable results emerge from this. For example, the free energy is
real at the complex saddle point and $\left\langle Tr_{F}P\right\rangle \ne\left\langle Tr_{F}P^{\dagger}\right\rangle $.
The mass matrix for Polyakov loops exhibits a new feature: the mass
eigenvalues may form a complex conjugate pair, indicating the occurrence
of spatially-modulated sinusoidal decay. Such behavior is forbidden
by spectral positivity for $\mu=0$, but is possible when $\mu\ne 0$.
In the case of PNJL models, complex conjugate pairs occur in regions
around the first-order line that emerges from $T=0$, terminating
at a critical end-point. 

Here we address the generality of this phenomenon by showing similar
behavior in lattice QCD with static quarks in the strong-coupling
limit. We will use a transfer matrix formalism to determine the behavior
of Polyakov loop correlation functions as a function of spatial separation.
These results are exact for any gauge coupling in $1+1$ dimensions,
but also represent the leading-order result in the character expansion
in higher dimensions. As the chemical potential of the static quarks
is varied, we will show that there are large regions of parameter
space where the eigenvalues of the transfer matrix form complex conjugate
pairs, leading to damped oscillatory behavior of Polyakov loop correlation
functions. The boundary of such a region in parameter space is referred
to as a disorder line in condensed matter physics. The method used
is completely different from the saddle point technique employed in
\cite{Nishimura:2014rxa,Nishimura:2014kla}, applied to a very different
model, illustrating the generality of the behavior. Any reliable simulation
method for finite-density lattice QCD should be able to reproduce
our results, which thus can serve as a benchmark for the validation
of algorithms.

Section II describes the strong-coupling formalism underlying our calculation.
We give a graphical demonstration of the non-hermiticity of the correlation
function matrix in character space when the chemical potential is
nonzero. We also discuss the symmetries of the model, paying particular
attention to particle-hole symmetry. Section III explains how the transfer
matrix for Polyakov loops can be realized in the character basis in
a form suitable for numerical diagonalization. In section IV, we present
our results for the Polyakov loop spectrum. A final section gives
our conclusions.

\section{Strong-coupling formalism}

\subsection{Setup}

Strong-coupling expansions and character expansions are well-developed
methods for exploring the properties of lattice gauge theories \cite{Drouffe:1983fv}.
Strong-coupling expansions are typically expansions in inverse powers
of some coupling $g^{2}$ around $1/g^{2}=0$. Generally such expansions
have a finite radius of convergence in an $1/g^{2}$, and thus are
not directly relevant for the continuum limit of non-Abelian gauge
theories at $g^{2}=0$. Nevertheless, they have often given important
insight into mechanisms and critical behavior. Character expansions
are closely related to strong-coupling expansions, but have many advantages.
Consider the case of $SU(3)$ lattice gauge fields in $1+1$ dimensions
at some finite temperature. In the absence of non-gauge fields, \emph{i.e.},
the quenched approximation, this model is exactly solvable. The action
$S_{p}$ of a single plaquette $U_{p}$ can be expanded in character
expansion
\begin{equation}
e^{S_{p}\left[U_{p}\right]}=\sum_{R}d_{R}c_{R}\chi_{R}\left(U_{p}\right)
\end{equation}
where $\chi_{R}$ is the character of an irreducible representation
$R$ of the gauge group $G$, $d_{R}$ is the dimensionality of $R$,
and $c_{R}$ is a coefficient that depends on the parameters of the
gauge action $S_{p}$. The character expansion is an expansion in
the ratios $c_{R}/c_{0}$, where $c_{0}$ is the coefficient of the
trivial (identity) representation. A strong-coupling expansion in
$1/g^{2}$ may be obtained by expanding these ratios. We will be using
the character expansion in what follows.

Our principle observable is the trace of the Polyakov loop operator
$P_{x}$ in irreducible representations of $SU(3)$. The Polyakov
loop operator $P_{x}$ is the time-ordered product of the timelike
links starting at a given point $x$ and returning to that point due
to the periodic boundary conditions of finite-temperature lattice
gauge theories. The trace of $P_{x}$ in an irreducible representation
$R$ measures confinement for that representation: $\left\langle Tr_{R}P\right\rangle =\exp\left(-\beta F_{R}\right)$
, where $F_{R}$ is the free energy required to insert a static particle
in a representation $R$ into the system and $\beta$ is the inverse
of the temperature $T$. In a pure gauge theory, the trace in the
fundamental representation, $Tr_{F}P$, is an order parameter for
confinement. 

\begin{figure}
\includegraphics[width=5in]{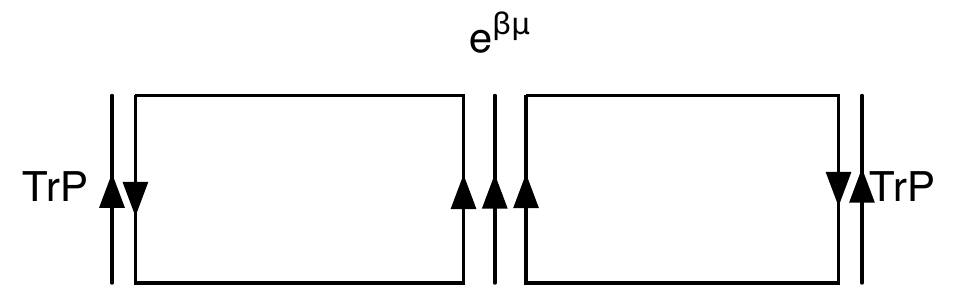}

\caption{\label{fig:TrP-TrP-1d}Graphical representation of a contribution
of fermions to $\left\langle Tr_{F}P\,Tr_{F}P\right\rangle $}.
\end{figure}

\begin{figure}
\includegraphics[width=5in]{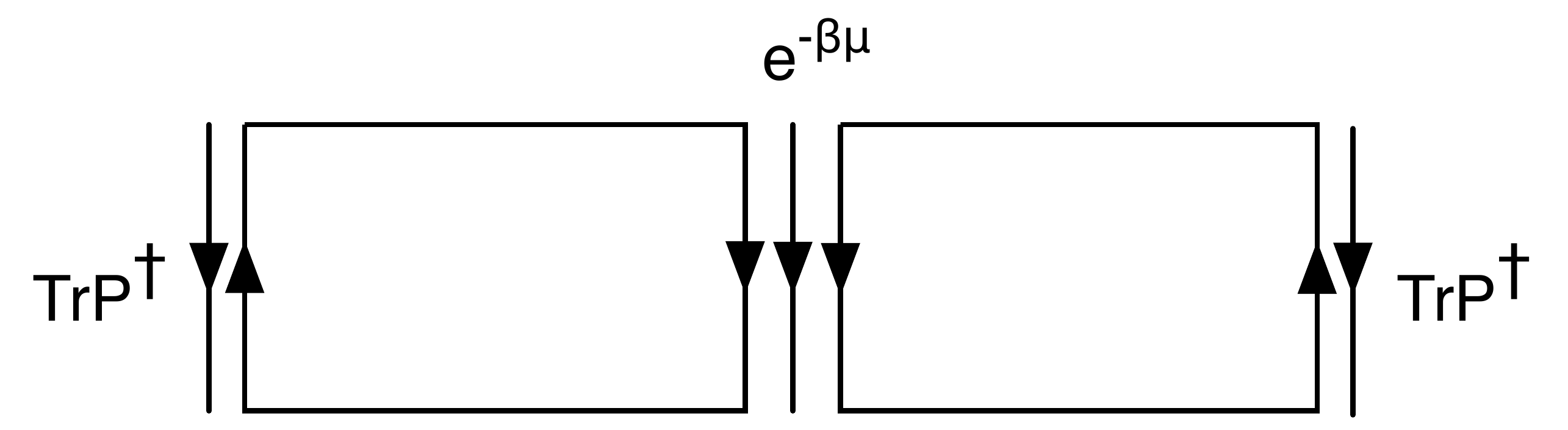}

\caption{\label{fig:TrPd-TrPd-2d}Graphical representation of a contribution
of fermions to $\left\langle Tr_{F}P^{\dagger}\,Tr_{F}P^{\dagger}\right\rangle $. }
\end{figure}

We will begin by giving simple arguments that show that the Polyakov
loop propagator matrix is not Hermitian at finite density, indicating the possibility
of complex eigenvalues. The correlation function of Polyakov loops
in a representation $R$, $\left\langle Tr_{R}P_{x}Tr_{R}P_{y}^{\dagger}\right\rangle $
is given to lowest order in the character expansion by tiling space
with plaquettes in the representation $R$ between $x$ and $y$.
We now introduce static quarks into this system. Each quark carries
with it a factor of $Tr_{F}P$ with an additional factor of $\exp(\beta\mu)$
when the chemical potential $\mu\ne 0$. As shown in Fig.~\ref{fig:TrP-TrP-1d},
this generates a new interaction not present in the quenched case
that couples $Tr_{F}P$ to itself. Correlation functions of the form
$\left\langle Tr_{R}P_{x}Tr_{R}P_{y}\right\rangle $ have been measured
in lattice simulation of full QCD at $\mu=0$ \cite{Doring:2007uh}.
When $\mu\ne0$, the lowest-order contribution to $\left\langle Tr_{R}P_{x}Tr_{R}P_{y}\right\rangle $
is enhanced by a weight factor $\exp(\beta\mu)$. On the other hand,
the coupling of $Tr_{F}P^{\dagger}$ to itself, as represented in
Fig.~\ref{fig:TrPd-TrPd-2d}, is suppressed by a corresponding factor
of $\exp-\beta\mu$. The only difference between the two graphs is
the factor of $\exp\left(\beta\mu\right)$ versus $\exp\left(-\beta\mu\right)$
so $\left\langle Tr_{R}P_{x}Tr_{R}P_{y}\right\rangle \neq\left\langle Tr_{R}P_{x}^{\dagger}Tr_{R}P_{y}^{\dagger}\right\rangle $. 
Thus when $\mu\ne0$ the matrix of two-point correlation functions
is no longer Hermitian. In order to find masses, that matrix of correlation
functions must be diagonalized. However, if the correlation matrix
is not Hermitian, the masses need not be real. The asymmetry between
$Tr_{F}P$ and $Tr_{F}P^{\dagger}$ is a consequence of the explicit
breaking of charge conjugation $\mathcal{C}$ by the chemical potential.
However, the symmetry of finite-density QCD under the combined action
of $\mathcal{C}$ and complex conjugation $\mathcal{K}$ remains intact
\cite{Ogilvie:2008zt,Ogilvie:2011mw,Nishimura:2014rxa,Nishimura:2014kla}.
Thus the combined action of $\mathcal{CK}$ leaves $Tr_{F}P$ and
$Tr_{F}P^{\dagger}$ invariant. For our purposes, the most important
consequence of the $\mathcal{CK}$ symmetry is that the masses are
either real or are part of a complex conjugate pair, as we discuss below.

A comprehensive treatment of Polyakov loop correlation functions at
strong coupling is most conveniently carried out using a
version of Svetitsky-Yaffe universality \cite{Svetitsky:1982gs,Billo:1996pu}.
This fundamental result for gauge theories at finite temperature links
the critical behavior of Polyakov loops in pure gauge theories in
$d$ dimensions with the behavior of spin systems in $d-1$ dimensions.
Furthermore, the effect of introducing quarks into the gauge theory
is similar to the effect of an external magnetic field on a spin system.
It is easy in strong coupling to show that the effective action for
the interaction of Polyakov loops is similar to that found in a spin
system. If we consider two adjacent Polyakov loops on a lattice, they
share a ``belt'' of plaquettes running up the time axis. Integrating
over the spatial links of this belt leads to an effective interaction
between the Polyakov loops of the form
\begin{equation}
\sum_{R}c_{R}^{N_{t}}\chi_{R}\left(P_{x}\right)\chi_{R}\left(P_{y}^{\dagger}\right)
\end{equation}
where $P_{x}$ and $P_{y}$ are Polyakov loops on adjacent spatial
lattice sites $x$ and $y$. The parameter $N_{t}$ is the temporal
size of the lattice, so $\beta=N_{t}a$, where $a$ is the lattice
spacing. To leading order in a strong-coupling expansion, the exponential
of the effective action is
\begin{equation}
e^{S_{eff}}=\prod_{\left\langle xy\right\rangle }\left[\sum_{R}c_{R}^{N_{t}}\chi_{R}\left(P_{x}\right)\chi_{R}\left(P_{y}^{\dagger}\right)\right]
\end{equation}
where the product is taken over all nearest-neighbor spatial lattice
sites. For very strong coupling, the contribution of the fundamental
representation usually dominates, and the effective action may be
written approximately as
\begin{equation}
e^{S_{eff}}\simeq\exp\left\{ \sum_{\left\langle xy\right\rangle }J\left[\chi_{F}\left(P_{x}\right)\chi_{F}\left(P_{y}^{\dagger}\right)+\chi_{F}\left(P_{x}^{\dagger}\right)\chi_{F}\left(P_{y}\right)\right]\right\} 
\end{equation}
where the sum over $\left\langle xy\right\rangle $ is a sum over
nearest-neighbor pairs of spatial points and
\begin{equation}
J=\left(\frac{c_{F}}{c_{0}}\right)^{N_{t}}.
\end{equation}
This is clearly of the form of a spin system, with spins taking on
values in $G$ and the interaction respecting global center symmetry.

In order to arrive at a simple model, we consider only the case where
quarks are so heavy that move only in time and are static in space.
In this case, the effects of quarks at $x$ can be represented in
the partition function by a weight \cite{Blum:1995cb}
\begin{equation}
D_{x}=\det\left[1+e^{\beta\mu-\beta M}P_{x}\right]
\end{equation}
while antiquarks give a weight factor
\begin{equation}
\bar{D}_{x}=\det\left[1+e^{-\beta\mu-\beta M}P_{x}^{\dagger}\right]
\end{equation}
where $\mu$ is the chemical potential and $M$ is the heavy quark
mass. It will be convenient to associate two different ``activities''
for quarks and antiquarks:
\begin{eqnarray}
z_{1} & = & e^{\beta\mu-\beta M}\\
z_{2} & = & e^{-\beta\mu-\beta M}.
\end{eqnarray}
Although $z_{1}$ and $z_{2}$ may take on any non-negative values,
their association with $\mu$ and $M$ does impose restrictions: Depending
on the sign of $\mu$,  either $z_{1}$ or $z_{2}$ is always
less than one. However, it is sometimes convenient to ignore this
restriction to display the symmetries of the model. The complete partition
function is given by
\begin{equation}
Z=\int\left[dP\right]\prod_{x}\left[D_{x}\bar{D}_{x}\right]\prod_{\left\langle xy\right\rangle }\left[\sum_{R}c_{R}^{N_{t}}\chi_{R}\left(P_{x}\right)\chi_{R}\left(P_{y}^{\dagger}\right)\right]
\end{equation}
where the integral $[dP]$ is over Haar measure for the Polyakov loop
$P_{x}$ on each spatial lattice site $x$ and the sum over $\left\langle xy\right\rangle $
is a sum over nearest-neighbor pairs.

\subsection{Symmetries}

The physics of $SU(N)$ static quarks at finite density is invariant
under $\left(z_{1},z_{2}\right)\rightarrow\left(z_{2},z_{1}\right)$
provided we also switch our definition of particle and antiparticle
operators. However, we need not switch operators at the special points
where $z_{1}=z_{2}$. These are the points where $\mu=0$ and we have
particle-antiparticle symmetry. Notice however that the identity 
\begin{equation}
\det\left[1+z_{1}P_{x}\right]=z_{1}^{N}\det\left[1+z_{1}^{-1}P_{x}^{\dagger}\right]
\end{equation}
also leads to an invariance and a symmetry. The factor of $z_{1}^{N}$
represents the Boltzmann factor for a completely filled state at a
site. Although this factor of $z_{1}^{N}$ does contribute to the
free energy, it does not affect expectation values. 
This operation reflects the equivalence between particles and holes:
a particle (relative to the vacuum) is equivalent to $N-1$ holes (relative
to the completely filled state) at the same site. In the special case
where $z_{2}=0$, there is an exact particle-hole symmetry under $z_{1}\rightarrow1/z_{1}$.
This extends to an approximate particle-hole symmetry when antiparticle
effects are small.
If we apply the above identity to antiparticles as well as particles, 
we obtain an exact symmetry of the complete theory:
The model is invariant under the transformation
$\left(z_{1},z_{2}\right)\rightarrow\left(1/z_{2},1/z_{1}\right)$.
Note however that  this transformation takes the physical
region $z_1, z_2 < 1$ into the unphysical region $z_1, z_2 >1$.

Let us suppose we are in a low-temperature, large $\mu$ region where
$z_{2}\ll1$ and antiparticle effects can be neglected. Then the fermion
determinant $D$ is real and there is no sign problem when $z_{1}=1$; 
this is precisely the absence of a sign problem at ``half-filling''
for static quarks. See \cite{Rindlisbacher:2015pea} for an extensive
treatment of this property. These symmetries are easily extended to
any representation of the gauge group. Note that an alternative approach
to including static quarks at finite density is to add a term
\begin{equation}
\sum_{x}\left[z_{1}\mbox{Tr}_F P_{x}+z_{2}\mbox{Tr}_F P_{x}^{\dagger}\right]
\end{equation}
directly to the action. However, this is an approximation to lowest
order in $z_{1}$ and $z_{2}$ of the effects of static fermions or
bosons. It therefore misses the effects of Pauli blocking as well
as the symmetries of the fermion determinant just discussed.

In addition to the invariances associated with $z_{1}$ and $z_{2}$, this
model inherits from finite-density QCD invariance under the combined
action of charge conjugation $\mathcal{C}$ and complex conjugation
$\mathcal{K}$  \cite{Nishimura:2014rxa,Nishimura:2014kla}. 
Charge conjugation takes $Tr_F P_{x}\rightarrow Tr_F P_{x}^{\dagger}$.
It is a symmetry when $\mu=0$, but not when $\mu\ne0$. Complex
conjugation is an antilinear symmetry, changing not only fields but
also complex-conjugating ordinary numbers. Like $\mathcal{C}$, $\mathcal{K}$
is a symmetry of the model only when $\mu=0$. $\mathcal{K}$ acts
on $Tr_F P$ to give $Tr_F P^{*}=Tr_F P^{\dagger}$. It is easy to see that
the combined effect of $\mathcal{CK}$ is to leave the action invariant.
From this, it can be shown that all the eigenvalues of the transfer matrix
are either real or are part of a complex conjugate pair. Due to the
symmetry of the model, the transfer matrix $T_s$ commutes with $\mathcal{CK}$.
If $T_s \left|\lambda\right\rangle =\lambda\left|\lambda\right\rangle $
it follows that
\begin{equation}
T_s\mathcal{CK}\left|\lambda\right\rangle =\mathcal{CK}T_s \left|\lambda\right\rangle =\mathcal{CK}\lambda\left|\lambda\right\rangle =\lambda^{*}\mathcal{CK}\left|\lambda\right\rangle 
\end{equation}
so $\lambda^{*}$ is an eigenvalue of $T_s$ if $\lambda$ is. Correlation
functions of operators that couple to eigenstates of $T_s$ with complex
eigenvalues will generally exhibit some amount of sinusoidally-modulated
exponential decay rather than the usual exponential decay found in
models with Hermitian actions \cite{Ogilvie:2008zt,Ogilvie:2011mw,Nishimura:2014rxa,Nishimura:2014kla}.
We will show below that this sinusoidal modulation is present in strong-coupling
QCD with a finite density of static quarks.

\section{Strong-coupling calculation of the spectrum}

In $1+1$ dimensions, the transfer matrix connecting one Polyakov
loop $P_{x}$ to its nearest neighbor $P_{y}$ in a pure gauge theory
can be written as
\begin{equation}
T_{0}=\sum_{R}c_{R}^{N_{t}}\chi_{R}\left(P_{x}\right)\chi_{R}\left(P_{y}^{\dagger}\right)
\end{equation}
in the gauge-invariant basis where states are class functions of $P$:
\begin{equation}
\Psi\left(P\right)=\sum_{R}b_{R}\chi_{R}(P).
\end{equation}
We can regard $T_{0}$ as acting on wave functions $\Psi(P)$ or alternately
on an infinite vector of coefficients $b_{R}$. We refer to the latter
representation as the group character basis. We are free to choose
the lattice action as reflected by the coefficients $c_{R}$, provided
they have the correct behavior in the continuum limit. Although the
Wilson action is the most common lattice action, there is an infinite
class of lattice actions that lead to the same continuum limit. Because
we are interested in tracking the behavior of Polyakov loop correlation
functions in many representations, we will need to keep the higher-order
terms in the character expansion of the effective action. We will
use the heat kernel action, for which the coefficients are
\begin{equation}
c_{R}=\exp\left(-g^{2}C_{R}a^{2}/2\right)
\end{equation}
where $C_{R}$ is the quadratic Casimir invariant for $R$. This has
important advantages for us over the standard Wilson action. The expression
for $c_{R}$ is simple and easy to calculate. In addition, it yields
exactly the continuum results for string tensions for pure gauge theories
in $1+1$ dimensions. See \cite{Marinov:1979gm,Menotti:1981ry,Billo:1996pu}
for an explanation of other properties of the heat kernel action.

Using the identification $\beta=N_{t}a$, the transfer matrix $T_{0}$
has the form
\begin{equation}
T_{0}=\sum_{R}\exp\left(-\beta g^{2}aC_{R}/2\right)\chi_{R}\left(P_{x}\right)\chi_{R}\left(P_{y}^{\dagger}\right).
\end{equation}
In the group character basis the Casimir operator is diagonal. For
$SU(3)$, its eigenvalues are 
\begin{equation}
C(p,q)=\frac{(p+1)^{2}+(q+1)^{2}+(p+1)(q+1)}{3}-1
\end{equation}
where $(p,q)$ specify the particular irreducible representation of
$R$ of the gauge group. Here $p$ represents the number of columns
of one box and $q$ represents the number of columns of two boxes
in Young tableau. The transfer matrix in the group character basis
is 
\begin{equation}
T_{0}=e^{-\frac{g^{2}\beta a}{2}C(p_{1},q_{1})}\delta_{p_{1}p_{2}}\delta_{q_{1}q_{2}}.
\end{equation}
In the pure gauge theory, the eigenvalues of $T_{0}$ determine the
exponential decay of correlation functions:
\begin{equation}
\left\langle \mbox{Tr}_{R}P_{x}\mbox{Tr}_{R}P_{y}^{\dagger}\right\rangle \sim\exp\left(-C(p_{1},q_{1})\frac{g^{2}\beta a}{2}\left|x-y\right|\right).
\end{equation}
It is convenient to define the combination $g^{2}\beta/2$ to be $m_{0}$,
so that each representation $R$ is associated with a mass $m_{p,q}\equiv C\left(p,q\right)m_{0}$
in the pure gauge theory. The eigenvalues of the pure gauge theory
transfer matrix $T_{0}$ are given by
\begin{equation}
\lambda_{p,q}=e^{-\frac{g^{2}\beta a}{2}C(p_{1},q_{1})}
\end{equation}
so a mass $m_{p,q}$ can be extracted as
\begin{equation}
m_{p,q}=-\frac{1}{a}\log\left(\frac{\lambda_{p,q}}{\lambda_{0,0}}\right).
\end{equation}
In the more general case where static quarks are present, the eigenvalues
of the transfer matrix cannot be associated with a fixed group representation,
and the corresponding eigenvectors in character space are linear combinations
of group characters. In general, we simply number the eigenvalues
sequentially starting at zero, and define the mass by
\begin{equation}
m_{j}=-\frac{1}{a}\log\left(\left|\frac{\lambda_{j}}{\lambda_{0}}\right|\right)
\end{equation}
taking into account that the eigenvalues may be complex. Note that
$m_{0}$ is simply a convenient mass scale, and not the mass of the
ground state. If we consider $m_{j}/m_{0}$ in the limit where quark
effects vanish, we obtain the Casimir operator $C(p,q)$. It is only
in this sense that a given mass can be associated with a representation
in the general case. In the case where $\lambda_{j}$ is complex,
$\mathrm{Arg}\left(\lambda_{j}\right)$ determines the wavenumber
for the oscillations. In general, we define
\begin{equation}
\exp\left(-m_{j}a+ik_{j}a\right)=\frac{\lambda_{j}}{\lambda_{0}}
\end{equation}
so that $k_{j}a=Arg\left(\lambda_{j}/\lambda_{0}\right)$ determines
the period of oscillation for a given eigenvalue $j$. In all the
cases considered here, the ground state is unique and $\lambda_{0}$
is real and positive, so $k_{j}a=Arg\left(\lambda_{j}\right)$.
We will take the lattice spacing $a$ to be $1$, and treat $m_0$
as the fundamental parameter of the pure gauge theory.

In the group character basis, $\mathrm{Tr}_F P$ and $\mathrm{Tr}_F P^{\dagger}$act
as raising and lowering operators and can be expressed in terms of
Kronecker deltas 
\begin{equation}
\mathrm{Tr}_F P=\delta_{p_{1,}p_{2}}\delta_{q_{1},q_{2}-1}+\delta_{p_{1,}p_{2}-1}\delta_{q_{1},q_{2}+1}+\delta_{p_{1,}p_{2}+1}\delta_{q_{1},q_{2}}
\end{equation}
\begin{equation}
\mathrm{Tr}_F P^{\dagger}=\delta_{p_{1,}p_{2}-1}\delta_{q_{1},q_{2}}+\delta_{p_{1,}p_{2}+1}\delta_{q_{1},q_{2}-1}+\delta_{p_{1,}p_{2}}\delta_{q_{1},q_{2+1}}.
\end{equation}
The effect of heavy static quarks can in turn be represented in the
partition function by the fermion determinant 
\begin{equation}
D\left(z_{1}\right)=1+z_{1}\mbox{Tr}_F P+z_{1}^{2}\mbox{Tr}_F P^{\dagger}+z_{1}^{3}
\end{equation}
while the effect of antiquarks is represented by 
\begin{equation}
\bar{D}\left(z_{2}\right)=1+z_{2}\mbox{Tr}_F P^{\dagger}+z_{2}^{2}\mbox{Tr}_F P+z_{2}^{3}
\end{equation}
where $z_{1}$ and $z_{2}$ are the ``activities'' for quark and
antiquark defined above. 
The overall transfer matrix including the effect of quarks
and antiquarks can be written as 
\begin{equation}
T_s=T_{0}^{1/2}D\left(z_{1}\right)\bar{D}\left(z_{2}\right)T_{0}^{1/2}.
\end{equation}
This particular form is chosen so that $T_s$ is Hermitian when $z_{1}=z_{2}$.
While the transfer matrix corresponding to pure gauge fields, $T_{0}$,
is Hermitian, the final transfer matrix, $T_s$, that includes the effect
of heavy quarks and antiquarks is no longer Hermitian. $\mathrm{Tr}_F P$
and $\mathrm{Tr}_F P^{\dagger}$ connects between different representations
so they introduce off diagonal elements in the transfer matrix. Since
$\mathrm{Tr}_F P$ and $\mathrm{Tr}_F P^{\dagger}$are different for $\mu\ne0$,
the off-diagonal elements are no longer symmetric and the transfer
matrix is non-Hermitian.

\begin{figure}
\subfloat[ In one dimension, the only graph is a straight line connecting the
endpoints (open circles).]{\includegraphics[width=3in]{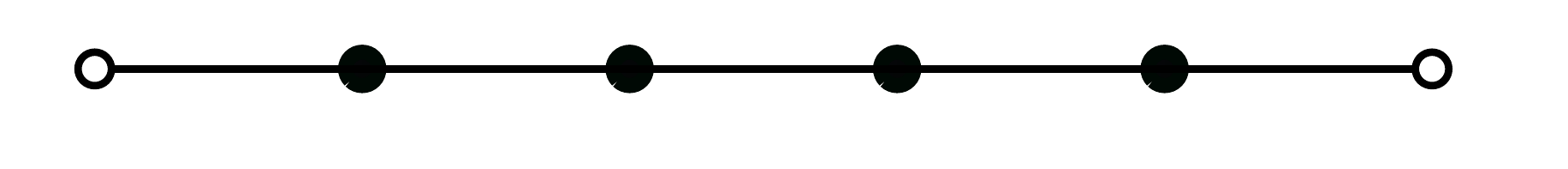}

}\subfloat[In two or more dimensions, off-axis correlation functions typically
have many paths of minimal length connecting the endpoints, but this
degeneracy does not change the rate of exponential fall-off at leading
order in strong coupling.]{\includegraphics[width=3in]{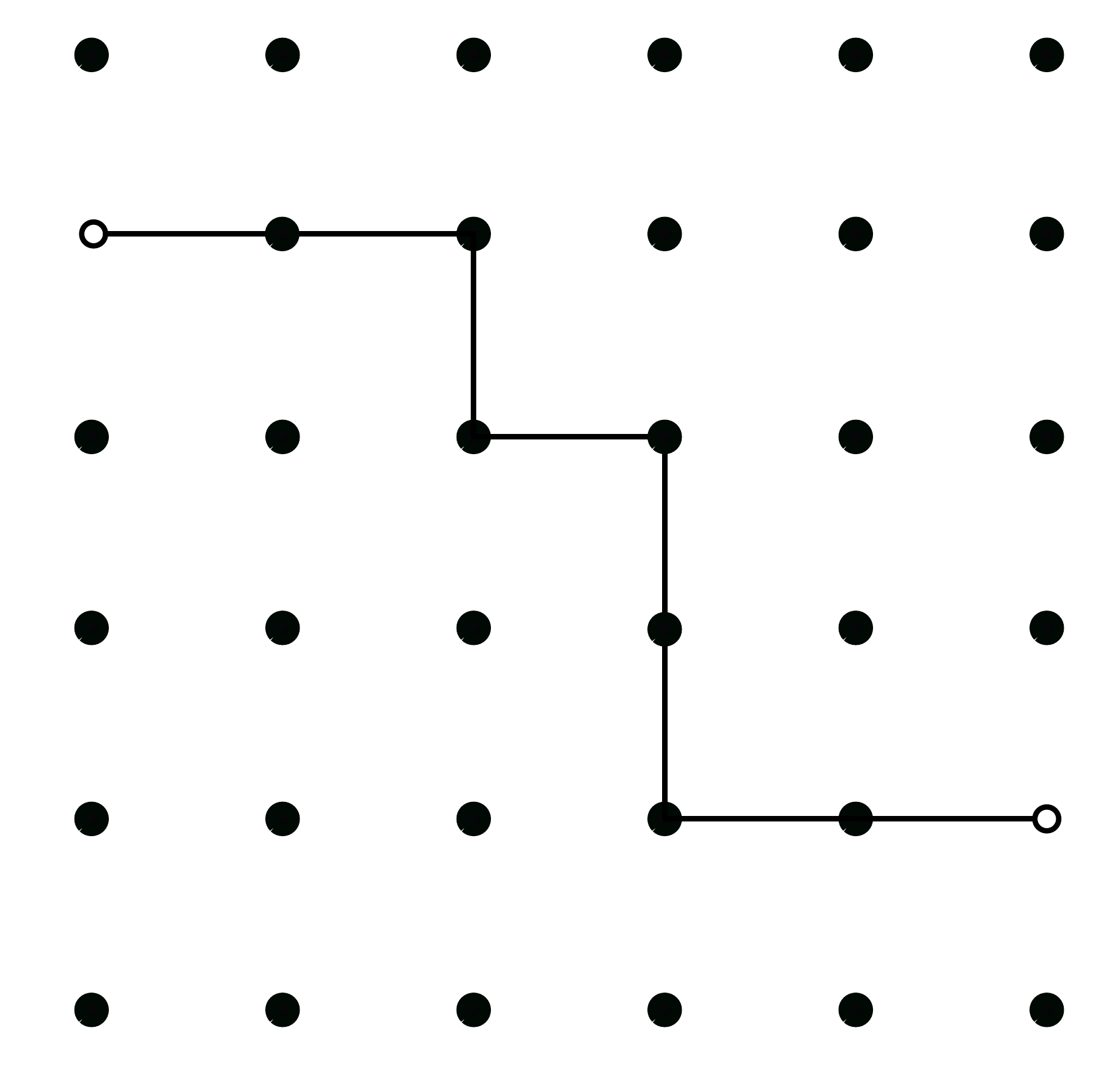}

}\caption{\label{fig:freescalar}Strong-coupling diagrams for a free scalar theory. }
\end{figure}

In $1+1$ dimensions, the results obtained from the transfer matrix
are exact at any value of the coupling. In the strong coupling region,
the results from $1+1$ dimensions are also valid in higher dimensions
to leading order in the character expansion. 
This result was noted long ago by \cite{Kogut:1981ny}.
Consider the strong-coupling expansion for a free scalar field, as shown
in Fig.~\ref{fig:freescalar}. 
For on-axis correlation functions, say
$\left\langle \phi\left(0\right)\phi\left(r\hat{x}\right)\right\rangle $,
the leading diagram is a line between the two points, the single path
of minimal length. For off-axis correlation functions, there are multiple
minimal-length paths in the taxicab metric,
e.g. $\left|x\right|+\left|y\right|+\left|z\right|$ in $3+1$ dimensions,
but this gives rise to a prefactor that does not change the exponential
decay of the correlation function. When looked upon as a spin system,
the diagrammatic expansion here behaves similarly, so the leading-order
strong-coupling result in $1+1$ dimensions is also the result in
$d+1$ dimensions. Higher-order corrections to masses do explicitly
depend on $d$. In any dimension, the strong-coupling expansion is
a convergent expansion with a finite radius of convergence, so there
will be some region around $g^{-2}=0$ where the lowest order result
is a good approximation. 

The analogous behavior for a gauge theory at finite temperature is
shown in Fig.~\ref{fig:3d-off-axis} below. For on-axis correlation
functions of widely-separated Polyakov loops, the dominant contribution
at strong coupling is a straight sheet, exactly as in $1+1$ dimensions.
For the off-axis correlation function, there will be many minimal
surfaces connecting the two Polyakov loops, but this will not change
the rate of exponential fall-off at leading order in strong coupling.
Thus we see that the strong-coupling results in $1+1$ dimensions
are also valid in $d+1$ dimensions.

\begin{figure}
\includegraphics[width=4in]{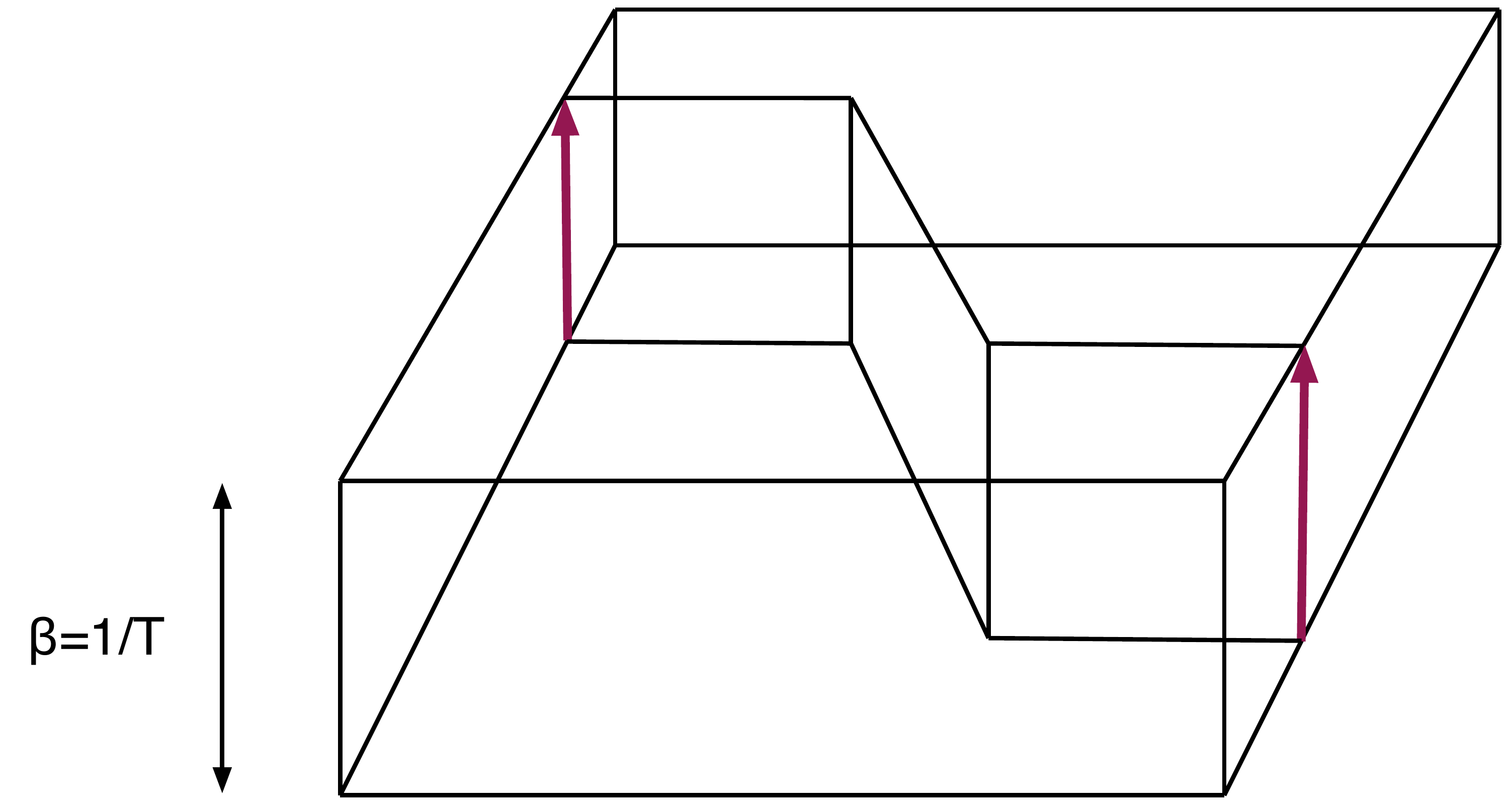}

\caption{\label{fig:3d-off-axis}A contribution to the off-axis correlation
function between two-widely separated Polyakov loops. For clarity,
the intermediate staircase has been replaced by a diagonal sheet.}

\end{figure}

\section{Results for the mass spectrum}

\begin{figure}
\includegraphics[width=5in]{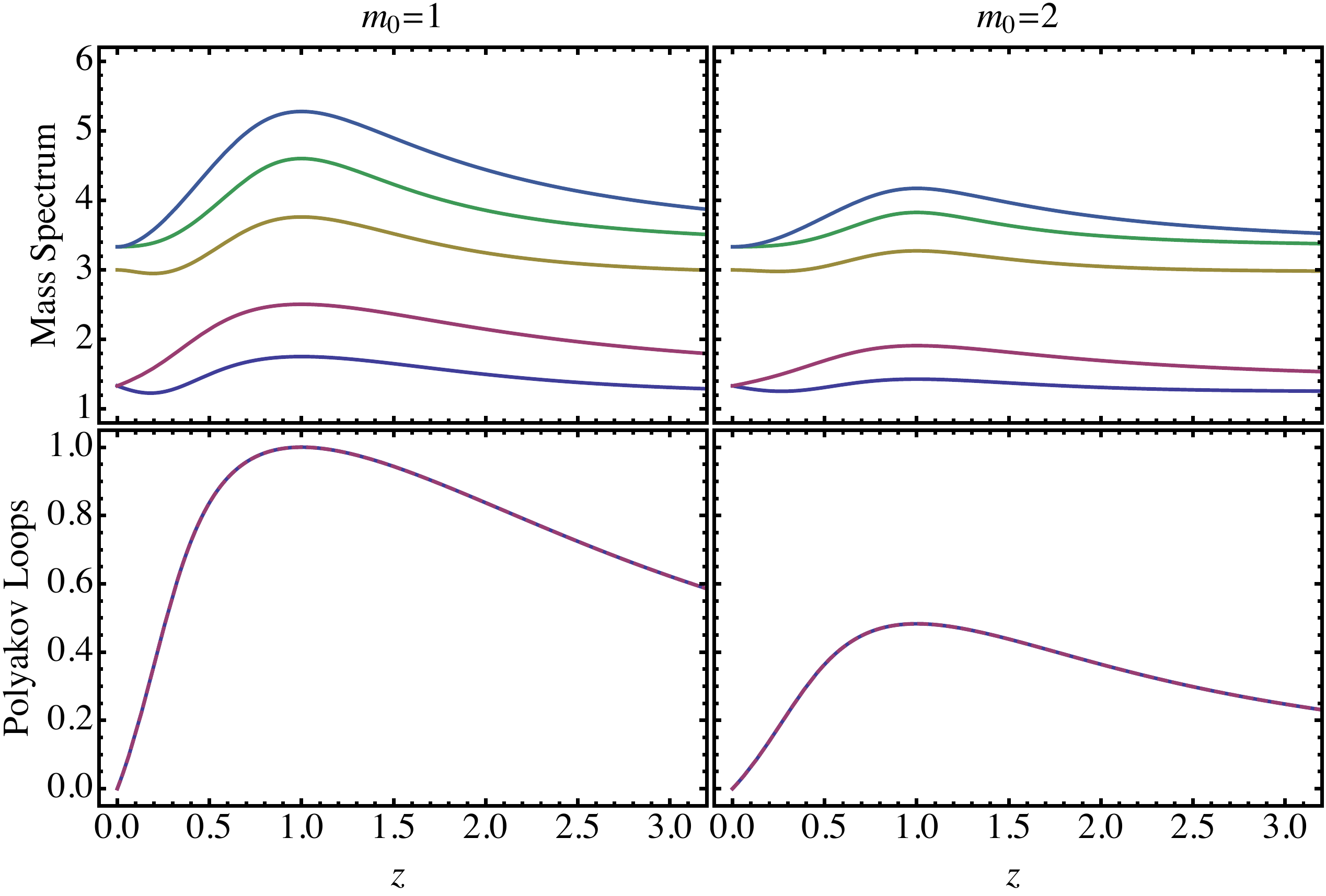}

\caption{\label{fig:Hermitian_m0=00003D1}The mass spectrum and Polyakov loop
expectation values for the $\left(1+1\right)$-dimensional $SU(3)$
model as a function of $z$ with $m_{0}=1$ and $m_{0}=2$. }

\end{figure}

We now discuss our results for the low-lying eigenvalues of $T_s$.
We begin with the case where $z_{1}$ and $z_{2}$ are set to a common
value $z$, which corresponds to setting $\mu$ to zero so that $z$
can be identified with $\exp(-\beta M)$. In this case, the model
is Hermitian so the eigenvalues are real
and $\left\langle Tr_{F}P\right\rangle =\left\langle Tr_{F}P^{\dagger}\right\rangle$. 
Figure \ref{fig:Hermitian_m0=00003D1} shows the mass spectrum of
the low-lying eigenstates and Polyakov loop expectations values for
$m_{0}=1$ and $2$. The masses shown in the figures below are always divided by the mass
scale $m_{0}$; the values of $m_j/m_0$ at $z=0$, on the left-hand axis, are thus
the values of the Casimir operator for low-lying representations of
$SU(3).$ In particular, we can associate the masses shown in 
Fig.~\ref{fig:Hermitian_m0=00003D1} with the $3$, $\bar{3}$, $8$, $6$
and $\bar{6}$ representations of $SU(3)$ when $z=0$. As can be
seen in the figure, the values on the left-hand axis at $z=0$ are
precisely $4/3$, $3$ and $10/3$. For $z>0$, the corresponding
eigenvectors contain a mixture of the identity representation, the
above five representations and other higher-dimensional representations.
Both the spectrum and Polyakov loop expectation values are invariant
under $z\rightarrow1/z$. This is reflected in the peaks achieved
at $z=1$ by
both the masses and the Polyakov loop expectation values.
Physically, quark effects behave as an external magnetic field,
and this effect is strongest at $M=0$, corresponding to $z=1$.
Increasing $m_{0}$ from $1$
to $2$ shortens the correlation length in lattice units, and decreases
the effects of the fermions on the spectrum. 
The peak in the Polyakov
loop at $z=1$ is smaller at $m_{0}=2$ than at $m_{0}=1$ because
the interaction between nearest-neighbor Polyakov loops is smaller.
 At $z=0$,
pairs of complex representations like the $3$ and $\bar{3}$ have
degenerate masses. Because the quark determinant breaks the $Z(3)$
symmetry of the pure gauge theory, the eigenstates for $z>0$ do not
show any degeneracy, but separate into clear, well-defined levels.

\begin{figure}
\includegraphics[width=5in]{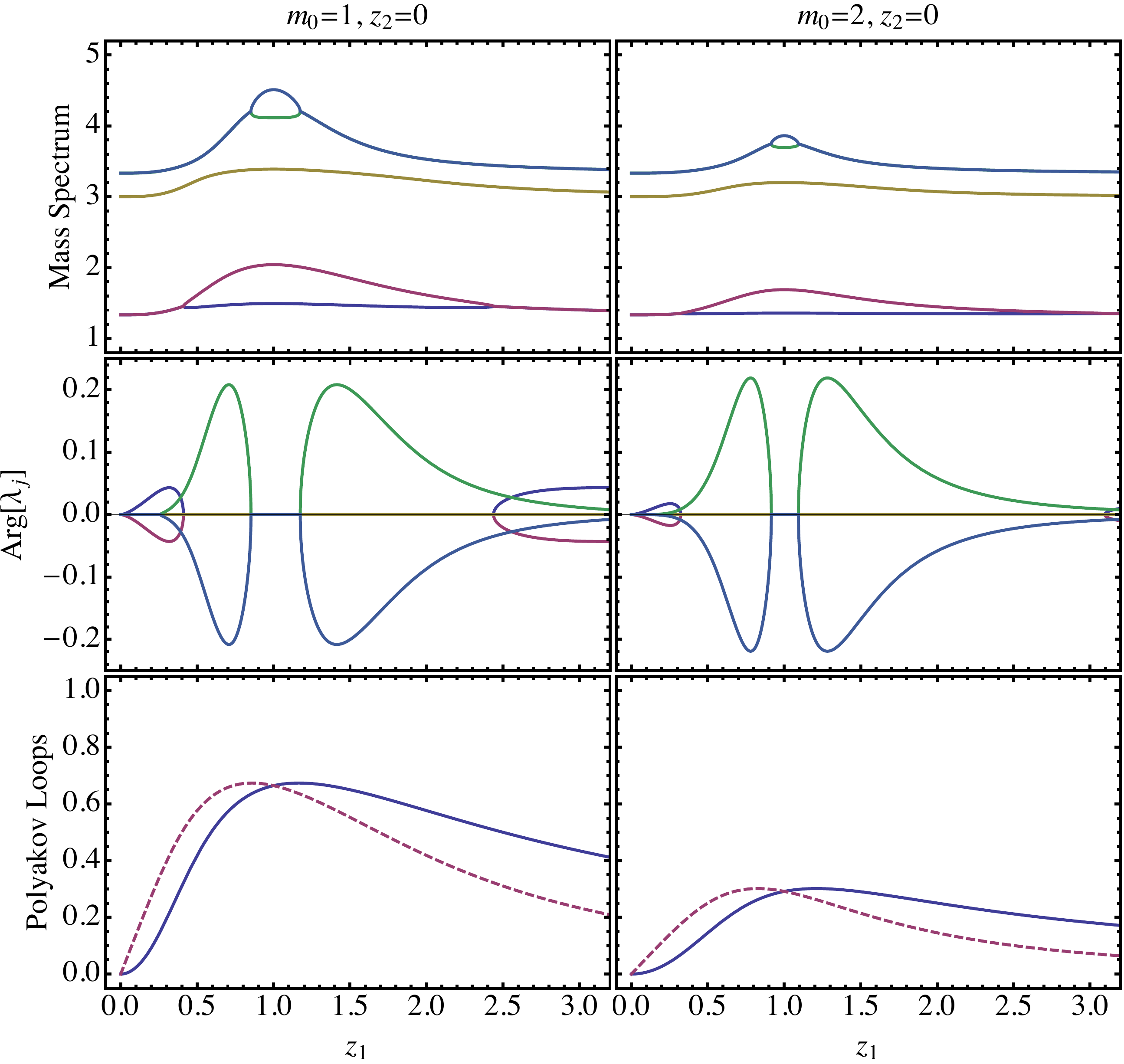}

\caption{\label{fig:z2_is_zero}The real and imaginary parts of the mass spectrum
and Polyakov loop expectation values for the $\left(1+1\right)$-dimensional
$SU(3)$ model as a function of $z_{1}$ with $m_{0}=1$ and $m_{0}=2$
and $z_{2}=0$. The Polyakov loop $\left\langle Tr_{F}P\right\rangle $
is represented by a solid line, and $\left\langle Tr_{F}P^{\dagger}\right\rangle $
by a dashed line.}
\end{figure}

We next consider the case where the effects of antiquarks
are neglected, which corresponds to setting $z_{2}=0$. 
Figure \ref{fig:z2_is_zero}
shows the real and imaginary parts of the mass spectrum,
$m_j/m_0$ and $Arg\left[ \lambda_j \right]$, for low-lying
eigenstates when $m_{0}=1$ and $2$. The figure also shows
Polyakov loop expectation values.
As seen in the plots, the masses start out real for $z_{1}=0$ but
quickly take on complex values for non-zero $z_{1}$. As $z_{1}$
increases, the magnitude of complex part of the mass gradually increases
before dropping back to zero. 
The plots clearly reflect the particle-hole symmetry
under $z_{1}\rightarrow1/z_{1}$. The real part of mass spectrum is
highest when $z_{1}=1$, which corresponds to $\mu=M$.
The point $z_{1}=1$ is special because
the theory is Hermitian at half-filling and the mass spectrum must be real.
Furthermore, there appears to be a region around $z_{1}=1$ where
the mass spectrum is real. The size of this region is largest for
the $3$ and $\bar{3}$ representations. The $8$ representation does
not develop an imaginary part. 
For $z_{1}<1$,
$\left\langle Tr_{F}P\right\rangle $ is less than $\left\langle Tr_{F}P^{\dagger}\right\rangle $,
implying that the free energy cost of inserting a fermion into the
system is greater than that of inserting an antifermion. For $z_{1}>1$,
this behavior is reversed in accordance with the $z_{1}\rightarrow1/z_{1}$
symmetry. At $z_{1}=1$, the two expectation values are equal.
As in the case $z_1=z_2$, moving from
$m_{0}=1$ to $m_{0}=2$ lessens the impact of fermions on the spectrum
and on Polyakov loop expectation values. As mentioned earlier, this
increase in scale of pure gauge theory lessens the effects of fermions
as seen in the plots. 

\begin{figure}
\includegraphics[width=5in]{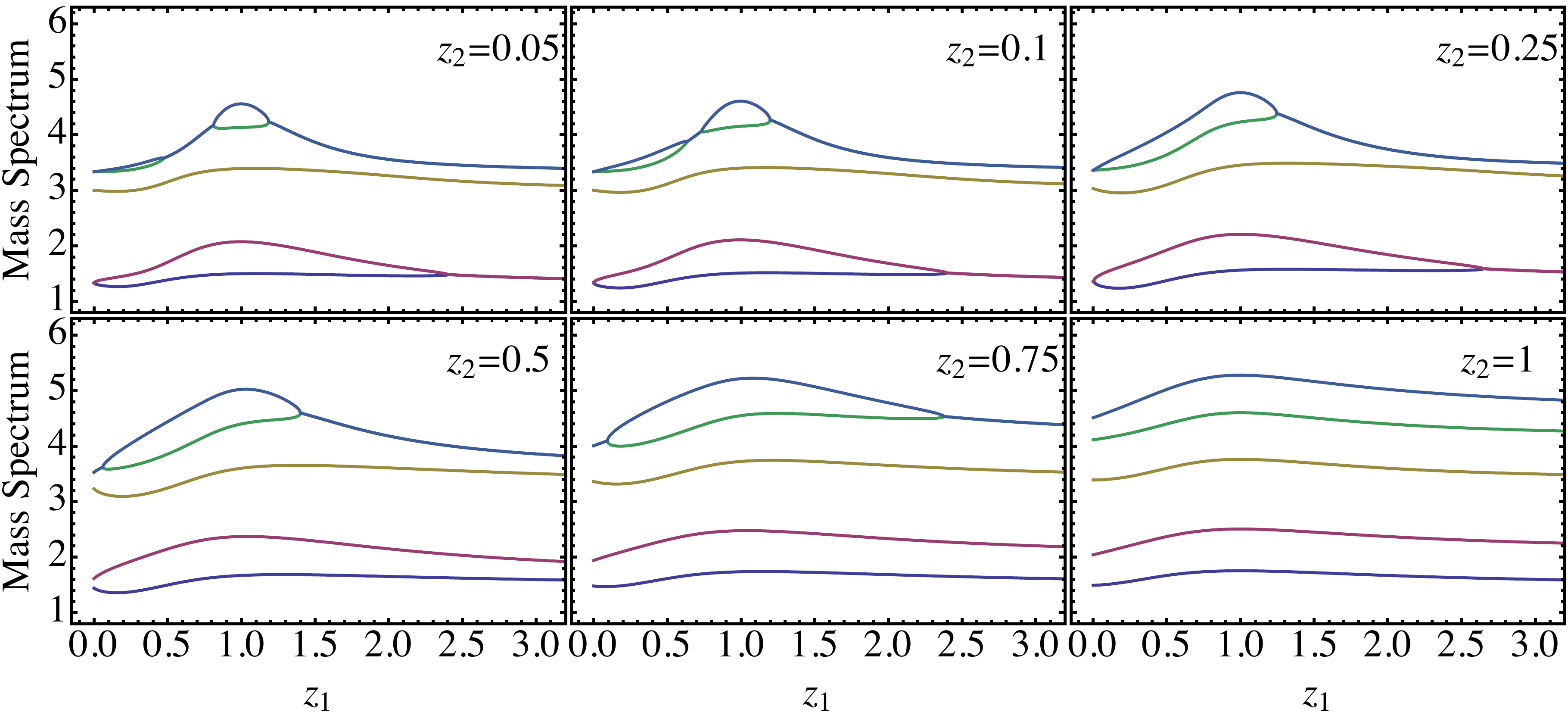}

\caption{\label{fig:turn_on_z2}The real part of the mass spectrum of the $\left(1+1\right)$-dimensional
$SU(3)$ model as a function of $z_{1}$ with $m_{0}=1$ and $z_{2}$
increasing.}
\end{figure}

We next consider the effects of antiquarks as gradually
``turned on'' by making $z_{2}$ non-zero. We fix the value of $m_{0}$
to $1$. 
The region where $Arg\left[\lambda_{j}\right]=0$ can be inferred
from the non-degeneracy of the real parts. As seen from Fig.~\ref{fig:turn_on_z2},
increasing the effect of antiparticles by making $z_{2}$ bigger gradually
shrinks the region of complex mass. The region of complex mass are
completely washed out for high enough value of $z_{2}$ and the mass
spectrum is completely real. The value of $z_{2}$ for which the eigenvalues
are real is smaller for low-lying eigenstates. For example, the $3$
and $\bar{3}$ eigenstates are real even when $z_{2}$ is around 0.5
but we need $z_{2}$ to be around 1 for the eigenstates corresponding
to $6$ and $\bar{6}$ to be completely real. When $z_{2}$ is much
larger than 1, the mass spectrum will again show regions where $Arg\left[\lambda_{j}\right]\ne0$.
This can be understood from the properties of the fermion determinants:
the antifermion determinant at large $z_{2}$ is equivalent to a fermion
determinant whose $z_{1}=1/z_{2}$ is small. The spectrum at large
$z_{2}$ is thus similar to that at small $z_{2}$.

\begin{figure}
\includegraphics[width=5in]{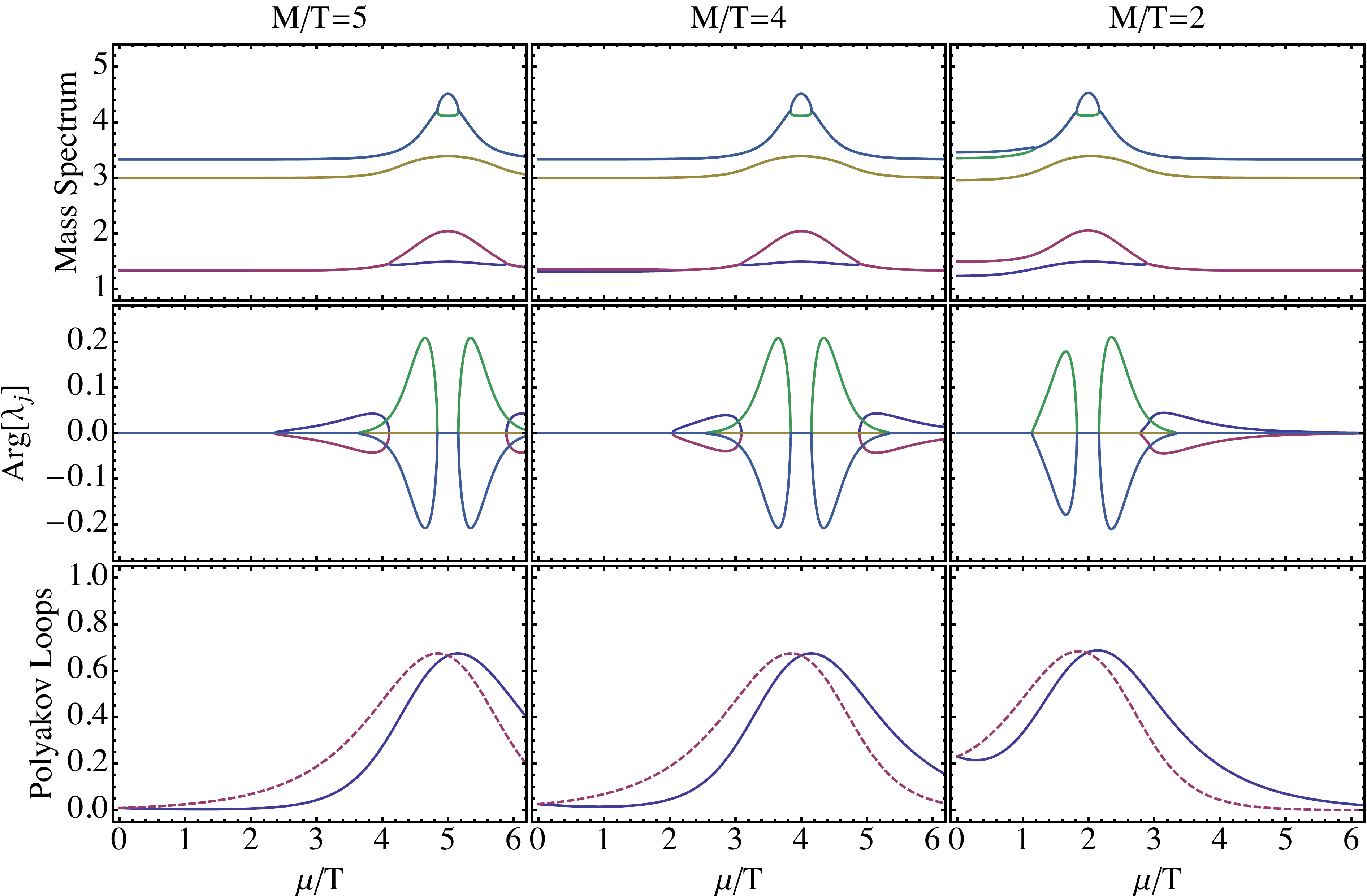}

\caption{\label{fig:mu_graphs_1}The real and imaginary parts of the mass spectrum
and Polyakov loop expectation values for the $\left(1+1\right)$-dimensional
$SU(3)$ model as a function of $\mu/T$ with $m_{0}=1$ and $M/T$
between $5$ and $2$. The Polyakov loop $\left\langle Tr_{F}P\right\rangle $
is represented by a solid line, and $\left\langle Tr_{F}P^{\dagger}\right\rangle $
by a dashed line.}
\end{figure}

\begin{figure}
\includegraphics[width=5in]{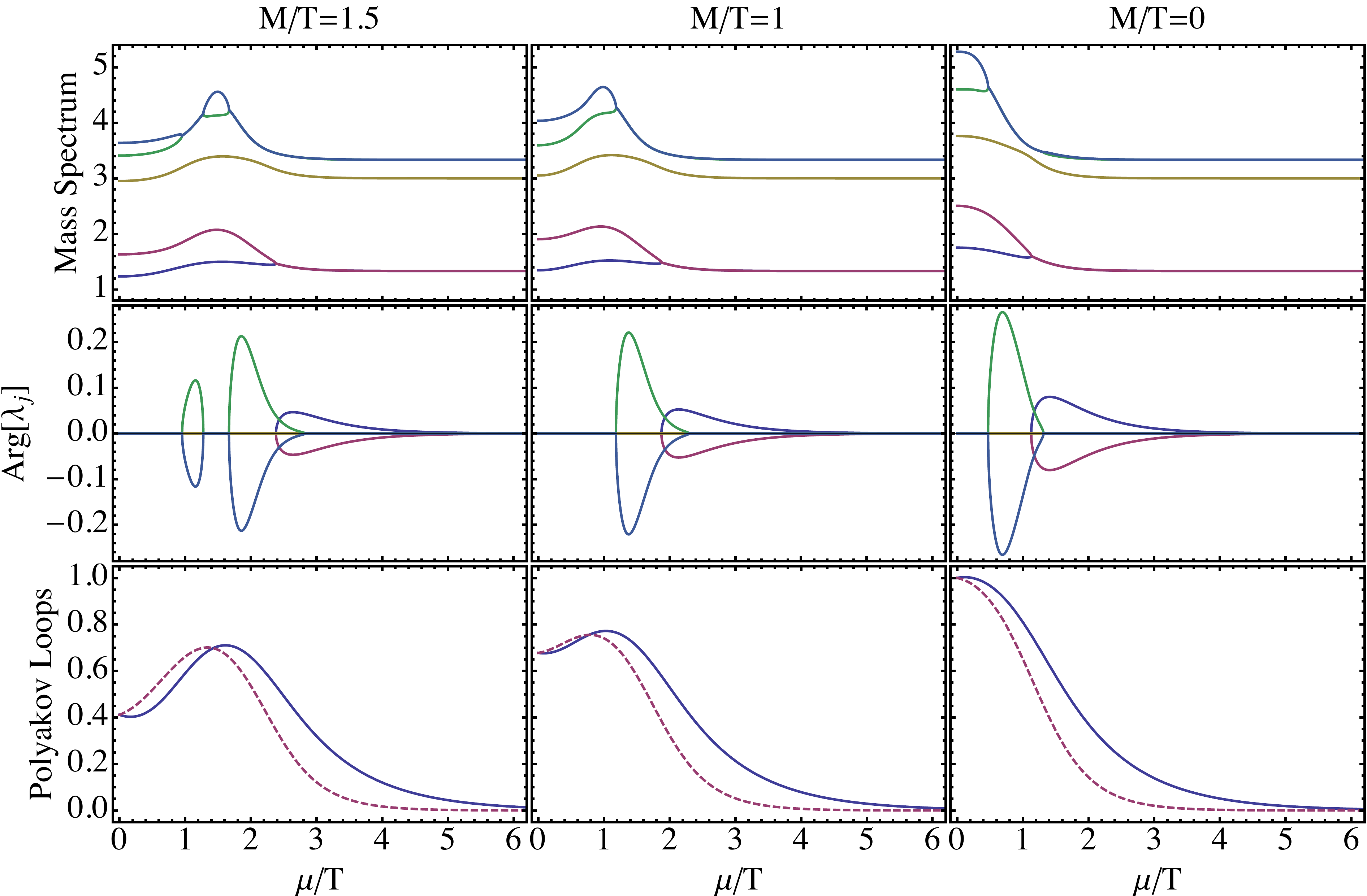}

\caption{\label{fig:mu_graphs_2}The real and imaginary parts of the mass spectrum
and Polyakov loop expectation values for the $\left(1+1\right)$-dimensional
$SU(3)$ model as a function of $\mu/T$ with $m_{0}=1$ and $M/T$
between $1.5$ and $0$. The Polyakov loop $\left\langle Tr_{F}P\right\rangle $
is represented by a solid line, and $\left\langle Tr_{F}P^{\dagger}\right\rangle $
by a dashed line. }
\end{figure}

We now turn to a more physical analysis of the spectrum in terms of
the quark mass $M$ and the chemical potential $\mu$. In all cases,
we set the fundamental scale-setting parameter $m_{0}=1$. The ratio
$M/T$ is fixed at values between $0$ to $5$ and $\mu/T$ is varied
from $0$ to $6$. As shown in Figs.~\ref{fig:mu_graphs_1} and \ref{fig:mu_graphs_2},
the behaviors of the mass spectrum and Polyakov loop expectation values
are similar to what has been seen before, but the peak in the real
part of the mass spectrum occurs near $\mu=M$, corresponding to $z_{1}=1$.
As before, the eigenvalue associated with the $8$ remains real throughout.
When $M$ is large compared to $m_{0}$ and $\mu$, the spectrum
is essentially that of the pure gauge theory. When $\mu$ is close
to $M$ there is again a clear region where the low-lying eigenstates
are real. In this region, there are well-defined eigenvalues associated
with the $3$ and $\bar{3}$, and with the $6$ and $\bar{6}$. 
Near $\mu=M$, there are also
clear maxima and minima in many of the mass values.
This is presumably due to a relative
maximum in the overall
strength of $Z(3)$ breaking at that point.
Furthermore, $Tr_F P$ and $Tr_F P^{\dagger}$ both
peak near $\mu=M$, and also cross near this point.
We attribute much of the observed behavior for $M \simeq \mu$
to an approximate particle-hole symmetry associated with
the transformation $z_{1}\leftrightarrow1/z_{1}$.
In the cases we are considering $z_{2}=\exp\left(-\beta\mu-\beta M\right)$
is typically less than $1$.
This leads to an approximate $z_{1}\leftrightarrow1/z_{1}$
symmetry with small corrections coming from $z_{2}<1$,
with the effects of the symmetry most pronounced in the region $M \simeq \mu$.
For $M/T$
large, $z_2$ is always much less than one, leading to 
an approximate particle-hole symmetry for all $\mu$;
crossing of the Polyakov loops occurs at approximately the same value of $\mu/T$
as half-filling.
For large $\mu$, Polyakov loop expectation values
go to zero, also as a consequence of the approximate symmetry.
Equivalently, one can view this as due to the saturation of the quark
number density at large $\mu$.
For $M/T \lesssim 0.55$, antiquark effects are significant and cannot be
neglected: the low-lying spectrum is complex at half-filling.

On either side of the region including
$M=\mu$, where the $3$ and $\bar{3}$ are real, there are regions
where the value of $\left|Arg\left[\lambda_{j}\right]\right|$ for
the $3$ and $\bar{3}$ representation reaches a maximum. These regions
also include the value of $\mu/T$ where $Tr_F P$ and $Tr_F P^{\dagger}$ are
most different. A similar correlation of $\left|Arg\left[\lambda_{j}\right]\right|$
with $\left|\left\langle Tr_{F}\left(P-P^{\dagger}\right)\right\rangle \right|$
was seen in PNJL models \cite{Nishimura:2014kla}. As in Fig.~\ref{fig:turn_on_z2},
there is a second region where the low-lying eigenvalues are all real.
This region is separate from the region around $M=\mu$ where the eigenvalues
are real, and appears here for low $\mu/T$. 
This behavior is clearly visible for $M/T=2$, but is present
for $M/T=4$ and even higher values.

\begin{figure}
\includegraphics[width=4in]{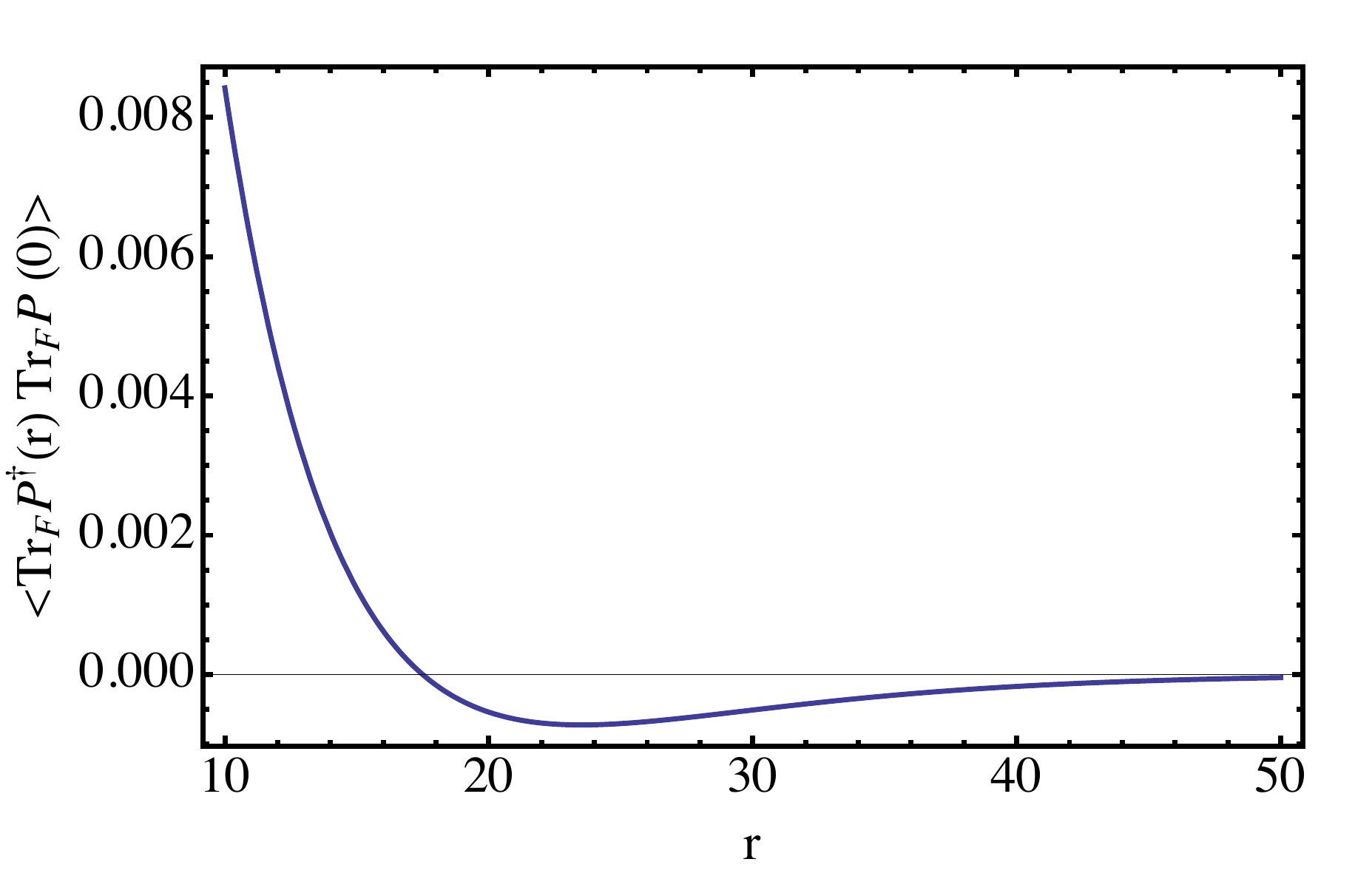}

\caption{\label{fig:corr_func}The $3-\bar{3}$ Polyakov loop correlation function
$\left\langle Tr_{F}P^{\dagger}\left(r\right)Tr_{F}P\left(0\right)\right\rangle $
as a function of $r$ for $m_{0}=0.1$ and $z_{1}=0.08$ with $z_{2}=0$.}

\end{figure}

As may be seen from Figs.~\ref{fig:mu_graphs_1} and \ref{fig:mu_graphs_2},
the magnitude of the imaginary part of any eigenvalue is generally
substantially smaller than the real part and may be difficult to observe
directly in simulations. 
Nevertheless, it may be possible
to observe the modulated decay directly in some circumstances. 
Figure \ref{fig:corr_func} shows the $3-\bar{3}$ Polyakov loop correlation
function $\left\langle Tr_{F}P^{\dagger}\left(r\right)Tr_{F}P\left(0\right)\right\rangle $
as a function of $r$ for $m_{0}=0.1$ and $z_{1}=0.08$ with $z_{2}=0$.
The clear minimum at $r\simeq22$ is a consequence of sinusoidal modulation
and reminiscent of the behavior of density-density correlation functions
in liquids. Note also that the correlation function drops below zero,
also as a consequence of the spatial modulation.

\begin{figure}
\includegraphics[width=4in]{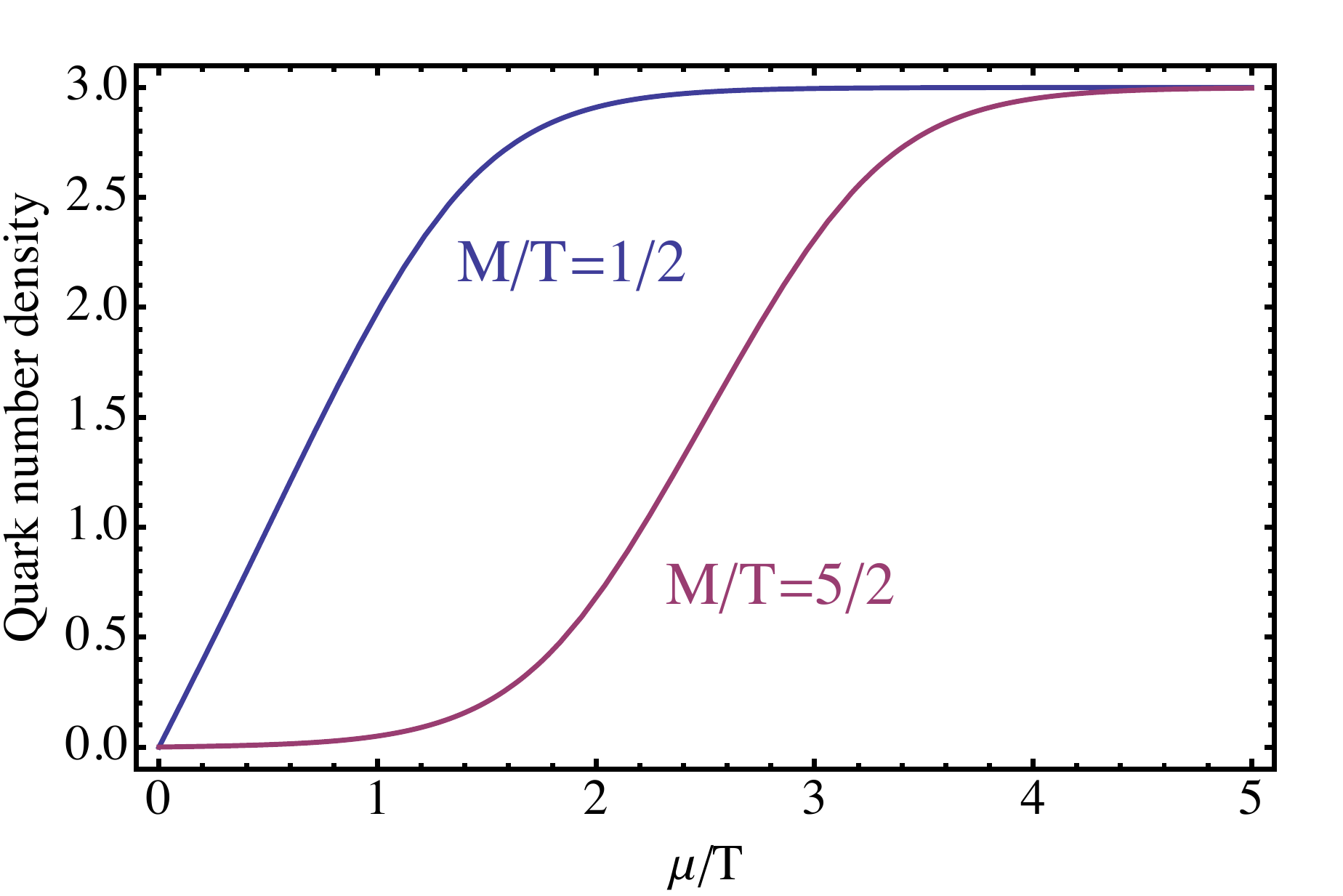}

\caption{\label{fig:Quark-number-density}Quark number density as a function
of $\mu/T$ for $M/T=1/2$ and $5/2$. The parameter $m_{0}$ is set
to $1$. For the heavier mass $M/T=5/2$, the quark number density
reaches half-filling $(1.5)$ at $M=\mu$. }

\end{figure}

The quark number density also may be calculated. Results are shown
in Fig.~\ref{fig:Quark-number-density} for the quark number density
as a function of $\mu/T$ for $M/T=1/2$ and $5/2$; the parameter
$m_{0}$ is set to $1$. The most obvious feature is the saturation
of the number density at $3$ for large $\mu$. In the context of
lattice gauge theories at finite density, saturation was first discussed
in \cite{Hands:2006ve} for the case of $SU(2)$, where there is no
sign problem. See \cite{Langelage:2014vpa,Rindlisbacher:2015pea}
for recent discussions of saturation effects in strong-coupling models
of $SU(3)$ at finite density. Saturation is also observed in recent
Langevin simulation with heavy quarks \cite{Aarts:2014bwa}. For the
heavier quark mass $M/T=5/2$, we expect that antiquark effects are
negligible for $\mu\simeq M$, and the system has an approximate particle-hole
symmetry at $M=\mu$. This in turn implies that the quark number density
reaches half-filling $(1.5)$ at $M=\mu$. As may be seen from the
figure, that expectation is confirmed. For the lighter quark mass,
$M/T=1/2$, antiquark effects are not negligible when $\mu\simeq M$,
and antiquark contributions lower the number density at $\mu=M$ below
the half-filling value. In both cases, the number density saturates
at some value of $\mu$ larger than $M.$ 

\section{Conclusions}

The spatial transfer matrix associated with Polyakov loops in finite-density 
QCD with static quarks have complex eigenvalues over a significant
region of parameter space in strong-coupling limit. The appearance
of complex eigenvalues is a direct consequence of the non-hermiticity
of the transfer matrix. This is a manifestation of the sign problem
in finite-density QCD. We have given a graphical explanation of the
non-hermiticity in terms of the mixing between different representations.
The invariance of finite-density QCD under $\mathcal{CK}$ symmetry
ensures that the eigenvalues are either real or part of complex conjugate
pair. The complex conjugate pairs in turn give rise to sinusoidal
modulation of Polyakov loop correlation function. 
If the activities $z_{1}$ and $z_{2}$ are set to a common value $z$, 
i.e. $\mu=0$, all the
eigenvalues are real. In this case, all the low-lying eigenvalues are largest
when the static fermion mass $M$ is zero, corresponding to $z=1$.
In the case where antiparticles effects are completely suppressed
by setting $z_{2}=0$, the mass spectrum reflected the 
particle-hole symmetry of
the theory under $z_{1}\rightarrow1/z_{1}$. As in the previous case,
the real part of the transfer matrix eigenvalues peak at $z_{1}=1$,
corresponding to $M=\mu$.
This is a point where the transfer matrix is Hermitian, and also
the point of half-filling.
The complete suppression of antiquark effects obtained by
setting $z_2=0$ is an approximation.
If $z_{2}$ were exactly zero, then we would find that: a) all mass
ratios would be real at $z_1=1$, a property that would hold
in a region around $z_{1}=1$; b) mass ratios would show
maxima or minima at $z_1 = 1$; c) $Tr_F P$ and $Tr_F P^{\dagger}$ would
cross at $z_1 = 1$. We have confirmed that these properties persist when
$z_2 \ll 1$, reflecting an approximate particle-hole symmetry.
As $z_{2}$ increases and antiparticle effects become more important,
the regions where complex eigenvalues occur reduce in size and eventually
disappear as $z_{2}$ approaches $z_{1}$. 

The mass spectrum can also
be analyzed in terms of the more physical parameters $M/T$ and $\mu/T$.
In general, the spectrum of low-lying eigenvalues shows a complicated
set of behaviors, with both real and complex pairs of eigenvalues
occurring. When $M/T$ is large compared to $m_{0}$ and $\mu/T$,
the spectrum obtained is essentially that of pure gauge theory, and
the low-lying eigenvalues are all real. 
The region where $\mu/T$ is close to $M/T$ corresponds to
$z_1 = 1$, and the behavior in this region is largely determined
by the approximate particle-hole symmetry discussed above.
In all the cases
studied, $\left\langle Tr_{F}P\right\rangle \le\left\langle Tr_{F}P^{\dagger}\right\rangle $
for $0<\mu<M$ reflecting the lower free energy cost associated with
introducing antiparticles. 
The occurrence of conjugate complex mass pairs and sinusoidal modulation
of Polyakov loop correlation functions was previously observed by
us in the study of phenomenological models of QCD using a saddle point
approximations \cite{Nishimura:2014rxa,Nishimura:2014kla}. The presence
of similar phenomenon in lattice models of QCD at strong coupling
strongly suggests the generality of the phenomenon. It has been suggested
\cite{Patel:2011dp,Patel:2012vn}
that sinusoidal modulation of this type might be observed in lattice
simulations and heavy ion experiments. In both phenomenological models
and in lattice strong-coupling calculations, the imaginary part of
the mass is significantly smaller than the real part, suggesting that
the direct observation of modulation might be difficult because of
a long wavelength. However, there is another way to observe the splitting
of the spectrum into complex pairs in lattice simulations. Suppose
that the imaginary part of the masses are too small to be directly
observed and can be neglected. In the regions where there are complex
conjugate eigenvalue pairs, the real parts of the eigenvalues are
degenerate, but outside of those regions they are different. This
effect should be present and observable in lattice simulations of
finite-density QCD, and may provide a strong test for finite-density
algorithms.
\begin{acknowledgments}
HN thanks Olaf Kaczmarek, Sandra Klevansky, Jan Pawlowski, Dirk Rischke,
Christian Schmidt and Ion Stamatescu for stimulating discussions.
The research of HN was supported by Bielefeld University and GSI.
\end{acknowledgments}

\bibliographystyle{unsrtnat}
\bibliography{spin_w_mu}
%\bibliography

\end{document}